\documentclass{article}
\usepackage{graphicx}
\usepackage{amsfonts}
\usepackage{mathtools}
 \usepackage{amssymb}
\usepackage[dvips]{color}
\date{\today}

\newcommand{\insertplot}[5]{\begin{figure}
 \hfill\hbox to 0.05in{\vbox to #5in{\vfill
 \inputplot{#1}{#4}{#5}}\hfill}
 \hfill\vspace{-.1in}
 \caption{#2}\label{#3}
 \end{figure}}
 \newcommand{\inputplot}[3]{
 \special{ps: plotfile #1}
}
\newcounter{fig}

\newcommand{\ee}{\end{equation}}
\newcommand{\eea}{\end{eqnarray}}
\newcommand{\be}{\begin{equation}}
\newcommand{\bea}{\begin{eqnarray}}

\begin{document}

 \title{ 
Strong gravity effects of charged $Q$-clouds and inflating black holes
} 

\author{{\large Yves Brihaye}$^{1}$, 
and {\large Betti Hartmann}$^{2,3}$
\\
\\
$^{1}${\small  Physique-Math\'ematique, Universit\'e de Mons-Hainaut, 7000 Mons, Belgium}\\
$^{2}${\small Instituto de F\'isica de S\~ao Carlos (IFSC), Universidade de S\~ao Paulo (USP), CP 369,13560-970 , S\~ao Carlos, SP, Brazil  }\\
$^{3}${\small Fakult\"at f\"ur Physik, Carl-von-Ossietzky Universit\"at Oldenburg, 26111 Oldenburg, Germany}}

\maketitle 
\begin{abstract} 
In this paper, we re-examine charged $Q$-clouds around spherically symmetric, static black holes. In particular, we demonstrate that for fixed coupling constants two 
different branches of charged scalar clouds exist around Schwarzschild black holes. This had not been noticed previously.
We find that the new solutions possess a ``hard wall'' at maximal possible gauge coupling. This wall separates the
interior (containing the black hole horizon), in which the scalar field is trapped in the ``false vacuum'', from the ``true vacuum'' exterior.
When taking back-reaction onto the space-time into account, we find that at maximal possible back reaction, the black hole solutions 
corresponding to these two branches either become extremal black holes with diverging scalar field derivative on the horizon or 
inflating black holes with a second, ``cosmological''  horizon which -- outside this second horizon -- correspond to extremal
Reissner-Nordstr\"om black holes. 
\end{abstract}

\section{Introduction}
Asymptotically flat black hole solutions of the (electro-)vacuum Einstein(-Maxwell) equations can be characterized by a very small number of parameters : their mass(, charge) and angular momentum and consequently uniqueness theorems for these type of black holes have been proven \cite{nohair}.
Moreover, a number of  No-hair theorems for scalar fields exist as well \cite{noscalarhair}, but these are not general. In fact, possibilities to
evade the requirements for these theorems exist and hence black holes can carry scalar hair.
An interesting observation in this respect is that when the scalar field possesses a harmonic time-dependence of the form
$\sim\exp(i\omega t)$, $\omega=\omega_R+i\omega_I$, a scalar wave impinging on a rotating black hole can be amplified if the real part of the frequency of the time-dependence $\omega_R$ is smaller than $m\Omega$, where $\Omega$
is the horizon velocity  and $m$ an integer \cite{amplification_scalar}. When, additionally, a bounding potential exists that traps the scalar modes, the black hole
can become super-radiantly unstable \cite{superradiance}. These phenomena have
led to the observation that when fulfilling the so-called ``synchronization condition''  $\omega_R=m\Omega$, $\omega_I=0$, Kerr black holes
with scalar hair exist \cite{scalar_Kerr}. Interestingly, these black holes possess a globally regular limit when letting the horizon tend to zero: so-called boson stars \cite{boson_stars}. These are objects made of massive scalar fields (that can be self-interacting) and can be interpreted
as macroscopic Bose-Einstein condensates. 
When considering static and charged black holes, i.e. Reissner-Nordstr\"om black holes, an equivalent condition exists for
scalar fields to be amplified. In this case, $\omega_R < g V(r_h)$, where $g$ is the U(1) charge of the scalar field and $V(r_h)$ the value
of the electric potential on the horizon \cite{RN_scalar}. However, a super-radiant instability of the RN solution is not possible \cite{no_superradiance_RN}. Fulfilling the ``synchronization condition'' $\omega_R= g V(r_h)$ leads to the existence
of charged black holes with scalar hair \cite{Hong:2019mcj, Herdeiro:2020xmb} (and also \cite{Hong:2020miv}).
In \cite{Hong:2019mcj}, the authors studied scalar clouds on Reissner-Nordstr\"om black holes, while
scalar clouds on Schwarzschild black holes as well as the back-reaction of the scalar clouds on the space-time have been addressed in \cite{Herdeiro:2020xmb}.  In particular, it was shown that the coupling to the electromagnetic field as well as self-interaction of the scalar
field are crucial ingredients for the existence of these clouds. 
In this paper, we re-examine and extend the results of \cite{Herdeiro:2020xmb}. In particular, we are interested
in charged clouds on Schwarzschild black holes and how these back-react on the space-time. 
We will show in the following, that two different types of clouds exist in the Schwarzschild background. 
Moreover, the back-reaction of these clouds leads to very different phenomena in the strong gravity regime. 
\section{The model}
In the following, we will discuss black hole solutions in General Relativity minimally coupled to an electromagnetic
and complex scalar field, respectively, in a $(3+1)$-dimensional space-time. The Lagrangian density reads~:
\begin{equation}
\label{eq:lag}
{\cal{L}} = \frac{{\cal R}}{16 \pi G} - D_{\mu} \Psi^{\dagger}  D_{\mu} \Psi - U(|\Psi|) - \frac{1}{4} F^{\mu \nu} F_{\mu \nu} \ ,
\end{equation}
where ${\cal R}$ is the Ricci scalar and $\Psi$ is a complex scalar field with potential $U(\vert\Psi\vert)$ that is minimally coupled to a $U(1)$ gauge field. $D_{\mu} \Psi= \partial_{\mu} \Psi - i g A_{\mu} \Psi$
and $F_{\mu \nu}=\partial_{\mu} A_{\nu} - \partial_{nu} A_{\mu}$ then denote the covariant derivative of the complex scalar field and field strength
tensor of the gauge field, respectively.  As show in \cite{Herdeiro:2020xmb}, $Q$-clouds exist
only for massive and self-interacting scalar fields. The potential we will use in the following is given by~:
\begin{equation}
\label{eq:pot_phi6}
U(\Psi)=\mu^2 \Psi^2 - \lambda \Psi^4 + \nu \Psi^6 \  .
\end{equation}
Due to appropriate rescalings (see below), we will be able to choose $\mu=\lambda=1$ without
loosing generality. The only parameter in the potential to vary will then be related to $\nu$. 
For $\mu=\lambda=1$, $\nu=1/4$, the potential
possesses three degenerate minima at  $\Psi=0$ , $\Psi^2 = 2$. 
On the other hand, for  $\mu=\lambda=1$, $\nu=1/3$, the potential has
a saddle point at $\Psi^2 =1$ and for $\nu > 1/3$ would be monotonically increasing. 
In our numerical construction, we will choose a value in between these two limiting values, see
more details below. 
We would like to  discuss static, spherically symmetric solutions to the equations resulting from
the variation of the action associated to (\ref{eq:lag}) and thence use the following Ansatz for the metric and matter fields
\begin{equation}
{\rm d}s^2 = -(\sigma(r))^2 N(r) {\rm d}t^2 + \frac{1}{N(r)} {\rm d}r^2 + r^2\left({\rm d}\theta^2 + \sin^2 \theta {\rm d}\varphi^2 \right) \ \ , \ \   A_{\mu} {\rm d} x^{\mu} = V(r) {\rm d} t  \ \ , \ \ 
\Psi=\psi(r) \exp(i\omega t) \ .
\end{equation} 
In the following, we will choose the gauge $\omega=0$, which due to the synchronization condition $\omega=g V(r_h)$ amounts to choosing $V(r_h)=0$. 
Using the rescalings $x=\mu r$, $v= \frac{\sqrt{\lambda}}{\mu} V$, $\psi = \frac{\sqrt{\lambda}}{\mu} \Psi$ we are left with three dimensionless couplings~:
\begin{equation}
\label{eq:couplings}
\alpha^2 = \frac{4\pi G \mu^2}{\lambda} \ \ , \ \   \beta^2 = \frac{\nu\mu^2}{\lambda^2} 
\ \ , \ \  e = \frac{g}{\sqrt{\lambda}}  \ .
\end{equation}
The equations of motion then read (the prime denotes the derivative with respect to $x$)~:
\begin{equation}
\label{eq:m_sigma}
  N'    = -2\alpha^2 x \left[ \frac{v'^2}{2 \sigma^2} + N \psi'^2 + U(\psi) + \frac{(e v \psi)^2}{N \sigma^2} \right]  - \frac{N - 1}{x} \  , \ 
                \sigma' =  2\alpha^2 x \sigma \left[ \psi'^2 + \frac{(ev \psi)^2}{N^2 \sigma^2} \right]  \ ,
\end{equation}
 \begin{equation}
 \label{eq:V}
      v'' = -  \left(\frac{2}{x} -\frac{\sigma'}{\sigma}\right) v' + \frac{2e^2 v \psi^2}{N}  \ ,
      \end{equation}
 \begin{equation}
 \label{eq:psi}
                        \psi'' = -\left(\frac{2}{x} + \frac{N'}{N} +\frac{\sigma'}{\sigma}\right) \psi' - \frac{e^2 v^2 \psi}{N^2 \sigma^2} + \frac{1}{2N}
                        \frac{dU}{d \psi}  \ .
\end{equation}
The boundary conditions for a black hole solution with regular matter fields on the event horizon at $x=x_h$ as well as finite energy and asymptotic flatness are~:
\begin{equation}
\label{eq:bc_horizon}
N(x_h)=0 \ \ , \ \   N'\psi'\vert_{x=x_h} = \frac{1}{2}\frac{dU}{d\psi}\biggr\vert_{\psi=\psi_h} \ \ , \ \  v(x_h)=0 
\end{equation}
with $\psi_h=\psi(x_h)$
and
\begin{equation}
\label{eq:bc_infinity}
N(x\rightarrow\infty)\rightarrow 1 \ \ , \ \  \psi(x\rightarrow\infty) \rightarrow 0 \ \ , \ \ 
v(x\rightarrow\infty)\rightarrow v_{\infty} \ .
\end{equation}
The physical parameters of the solutions are the (dimensionless) mass $M$, the (dimensionless) electric charge $Q$ and the
(dimensionless) Noether charge $Q_N$.  These can be read off from the metric and matter field functions at infinity
\begin{equation}
\label{eq:infty}
N(x \gg 1)=1-\frac{2M}{x} + \frac{\alpha^2 Q^2}{x^2} + ..... \ \ , \ \ 
v(x)=v_{\infty}  - \frac{Q}{x} + ....
\end{equation}
and the following integral of the $t$-component of the locally conserved Noether current~:
\begin{equation}
\label{eq:noether}
Q_N=\int {\rm d} x \ \frac{2x^2 e v\psi^2}{N\sigma}  \ .
\end{equation}
This globally conserved quantity can be interpreted as the number of scalar bosons and is related
to the electric charge $Q$. In fact, using the equation (\ref{eq:V}) for $v$, it is easy to show that for globally regular solutions $eQ_N\equiv Q$, i.e. the total charge is the charge of $Q_N$ individual
scalar bosons that each carry charge $e$. For black holes, this is no longer true, since
the horizon at $x=x_h$ presents a surface, on which regularity (boundary) conditions have to be imposed. The total electric charge is then a sum of the electric charge of $Q_N$ individual
charges $e$ and the horizon electric charge given by $-E_x(x_h) x_h^2/\sigma(x_h)$,  
where $E_x(x)=-v'(x)$  is the radial electric field.
The scalar field has an exponential decay at infinity, which reads
\begin{equation}
\label{eq:fall_off_scalar}
\psi(x\rightarrow \infty)\sim \frac{\exp(-\mu_{\rm eff,\infty} x)}{x} + .... \ \ , \  \
\mu_{{\rm eff},\infty}=\sqrt{1 - e^2 v_{\infty}^2} \ .
\end{equation}
Hence, the scalar field has an (asymptotic) effective mass given by the difference between the ``bare'' mass $\mu\equiv 1$ and the electric
potential energy. Obviously, the requirement of $e v_{\infty} < 1$ then is the condition that the electric potential energy
should be lower than the threshold to produce scalar particles of mass $\mu$.   
We can also define the temperature of the black hole, which will be important in the strong gravity regime.  For our choice
of metric this is given by
\begin{equation}
 T_H=(4\pi)^{-1}\sigma(x_h) N'\vert_{x=x_h}   \ .
\end{equation}
\section{$Q$-clouds around static black holes}
In the following, we assume $\alpha^2\equiv 0$, i.e. neglect the back-reaction of the scalar and gauge field,
respectively, on the space-time. There are two cases to discuss here, which turn out to be very different in the strong gravity limit. For Reissner-Nordstr\"om black holes, the electric field is completely fixed and does not interact dynamically with the scalar field.  We will hence discuss this case first. When choosing a Schwarzschild black hole background, the gauge field interacts
dynamically with the scalar field and we will show in the following that this has interesting
consequences, in particular when letting these charged $Q$-clouds back-react on the space-time. 
In the following, we will also be interested in the mass of the $Q$-cloud, $M_{Q}$ as defined in \cite{Herdeiro:2020xmb}~:
\begin{equation}
M_{Q}=\frac{1}{4\pi} \int {\rm d}^3 x \ \sqrt{-g} \left(T^i_{\ i} - T^t_{\ t}\right) \ \ , \ \  i=1,2,3 \ .
\end{equation}
Using the explicit expressions for the energy-momentum tensor and the gauge field
equation, we find that
\begin{equation}
\label{eq:mq}
M_Q = 2 v_{\infty} Q - \int\limits_{x_h}^{\infty} {\rm d} x \ \frac{v'^2 x^2}{\sigma} - 2 \int\limits_{x_h}^{\infty} {\rm d} x \ x^2 \sigma \ U(\psi)  \ .
\end{equation}
The second term on the (rhs) of this relation is the energy stored in the electric field of the cloud, since
the integrand is nothing else but $F_{\mu\nu} F^{\mu\nu} \sqrt{-g} \ {\rm d} x$. 
Hence, the first two terms are related to the electromagnetic field.
The part of the ``bare'' scalar field that gravitates is the scalar potential energy only.
\subsection{Reissner-Nordstr\"om black holes}
Rescaling $v\rightarrow v/\alpha$ and $e\rightarrow \alpha e$ and letting $\alpha=0$ leads to the decoupling of the 
scalar field equation from the remaining equations \cite{Herdeiro:2020xmb}. The gravity and electromagnetic field equations have a well known solution, the Reissner-Nordstr\"om (RN) solution, which is a non-extremal (extremal, respectively) black hole solution for charge $Q$ smaller (equal) to the mass $M$ and in this case reads~:
\begin{equation}
  v(r) = \frac{Q}{x_h} - \frac{Q}{x} \ \ ,  \ \ \sigma\equiv 1  \ \ , \ \    N(x) = 1 - \frac{2M}{x} + \frac{Q^2}{x^2} \ .
\end{equation}
$x_h$ denotes the radius of the outer (event) horizon of this black hole given by $x_h=M+\sqrt{M^2- Q^2}$, which becomes $x_h^{\rm (ex)}=M=Q$ for the extremal case $M=Q$.  Note that here $v_{\infty}=Q/x_h$ is fixed by the space-time.  We then consider the scalar field equation in the background of this black hole solution.
The scalar field equation in the presence of the fixed metric and electromagnetic fields can be interpreted as that of a scalar
field in an effective potential $U_{\rm eff}=\mu_{\rm eff}^2 \psi^2 - \psi^4 + \beta^2 \psi^6$, where 
\begin{equation}
\mu_{\rm eff}^2 = 1 - \frac{e^2 v^2}{N} = 1- e^2 \frac{Q^2 (x_h-x)}{x_h (Q^2 - x x_h)}  
\end{equation}
is the effective ``mass'' of the scalar field, which is $\mu_{\rm eff}^2 = 1 - \epsilon Q^2/(x_h^2-Q^2) + {\cal O}(\epsilon^2)$ close
to the horizon at $x=x_h+\epsilon$ and asymptotically becomes $\mu_{{\rm eff},\infty}$ (see (\ref{eq:fall_off_scalar})). 
The scalar field equation then has to be solved numerically subject to the  regularity condition on the horizon (see (\ref{eq:bc_horizon}))~:
\begin{equation}
\label{eq:horizon_condition}
   \psi'\vert_{x=x_h}  = \frac{x_h^3}{2(x_h^2-Q^2)} \frac{dU}{d\psi}\biggr\vert_{\psi=\psi_h} 
       \ \ \ .
\end{equation}
where $\psi_h$ denotes the value of the scalar field on the horizon, $\psi(x=x_h)$. 
The condition (\ref{eq:horizon_condition}) tells us that the value of the scalar field on the black hole horizon, $\psi_h$, has to
be  fine-tuned (using the gauge coupling  $e$) such that 
the scalar field  decays exponentially according to (\ref{eq:fall_off_scalar}).
This case has been studied in detail in \cite{Hong:2019mcj}, so we refer the reader for more details
to this paper. Here, we have cross-checked our numerics with that in the aforementioned paper
and find perfect agreement. 

\vspace{1cm}
\begin{figure}[ht!]
\hspace{-1cm}
\includegraphics[width=8.5cm]{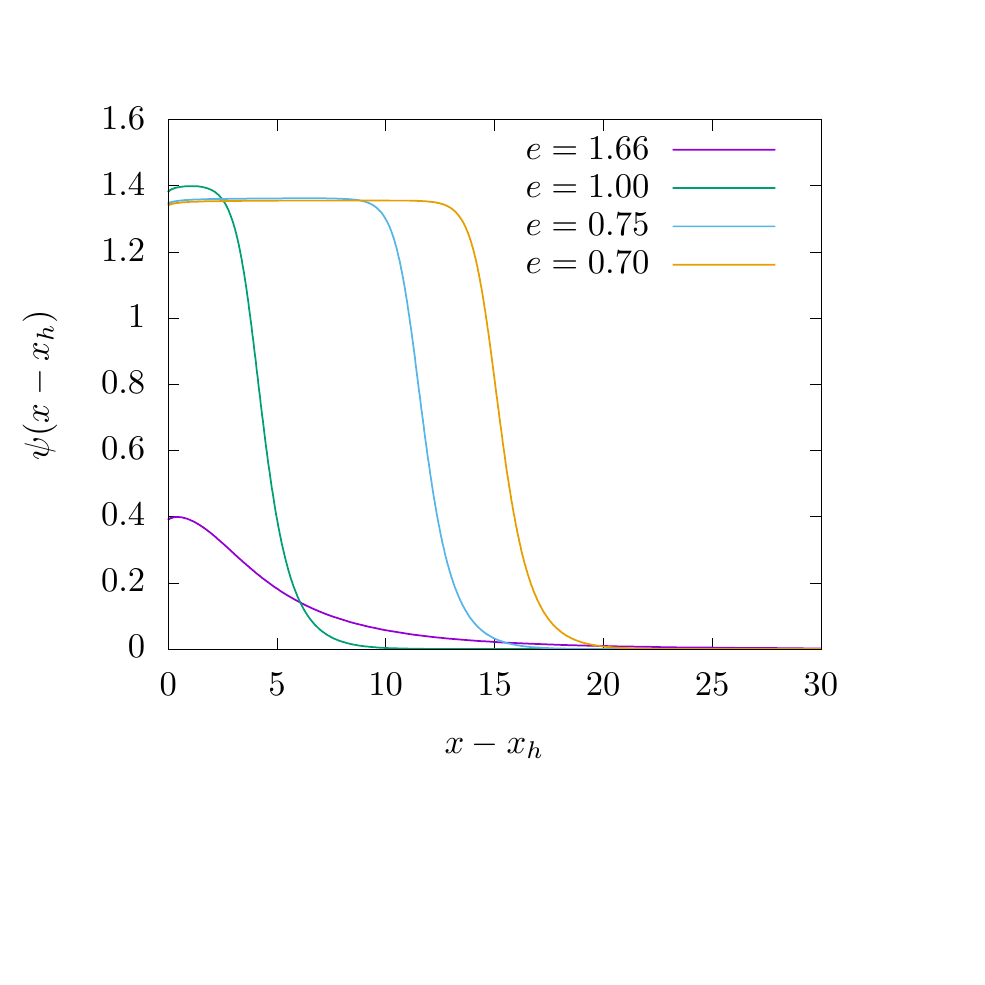}
\includegraphics[width=8.5cm]{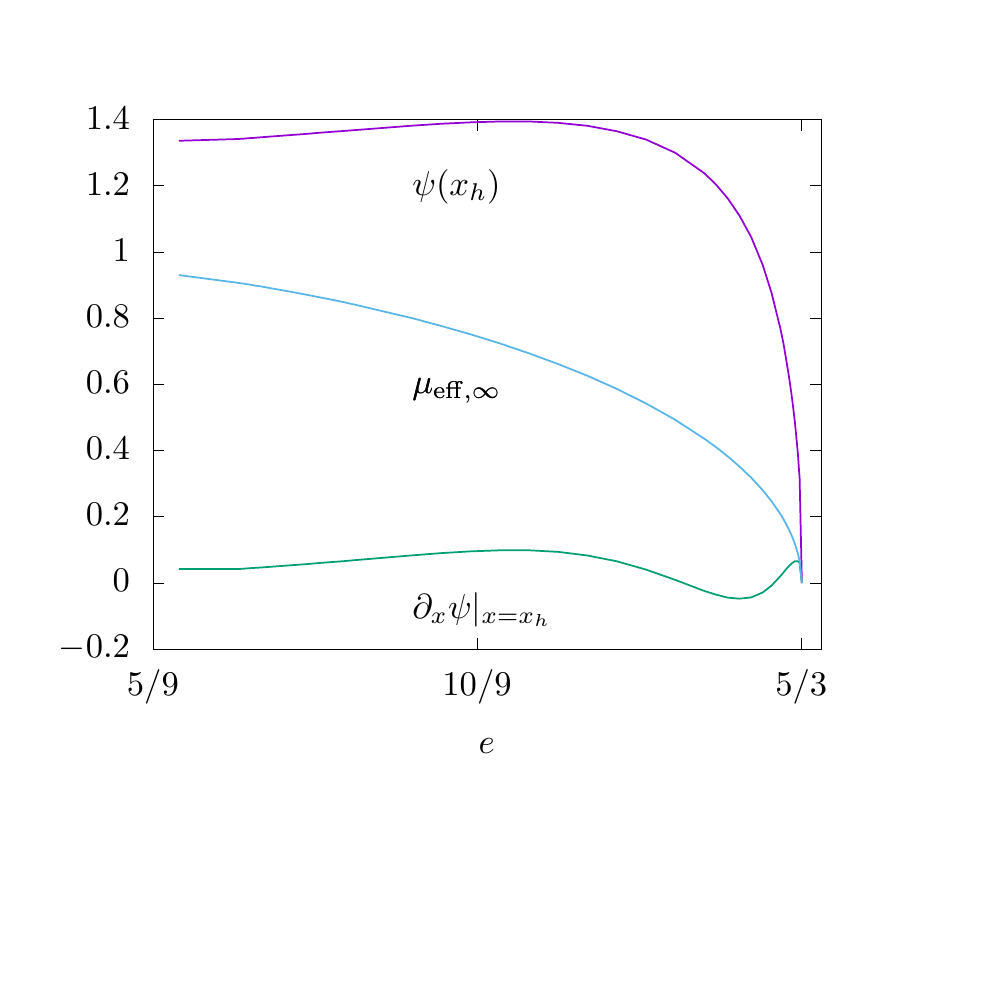}
\vspace{-2.5cm}
\caption{{\it Left}~: We show the profile of the scalar field function $\psi(x)$ for scalar clouds on a RN black hole with horizon radius $x_h=0.15$ and charge $Q=0.09$ 
for different values of the gauge coupling $e$. {\it Right}: The values of the scalar field, $\psi$ and its derivative $\partial_x \psi\equiv \psi'$ on the horizon as well
as $\mu_{{\rm eff},\infty}$ in function of $e$ for scalar clouds on this same RN black hole.   }
\label{data_rn}
\end{figure}

We have chosen $\beta^2 = 9/32$ throughout this paper.  As (\ref{eq:fall_off_scalar}) demonstrates, scalar clouds  can only exist close to a
RN black hole if $e < e_{\rm max}=x_h/Q$. Moreover,  the coupling of the scalar
field to the electromagnetic field needs to be sufficiently strong, i.e. $e \geq e_{\rm min}\neq 0$. In fact, decreasing $e$ from its maximal possible value $x_h/Q$, the cloud becomes increasingly extended. We find numerically that it extends to arbitrarily large
$x$ for $e\rightarrow e_{\rm min}$. The reason for this is that the second local minimum of the effective potential $U_{\rm eff}$ 
located at $\psi_{{\rm min},2}=(1+\sqrt{1-3\beta^2 \mu_{\rm eff}^2})/(3\beta^2)$ becomes degenerate in its potential value with the local
minimum  $\psi_{{\rm min},1}=0$ for $\mu_{\rm eff}=1/(2\beta)$.  This imposes a constraint on how strong the 
scalar field can be coupled to the electromagnetic field of the black hole. Using $\mu_{\rm eff}^2\approx 1-e^2 Q^2/x_h^2$ we find that
\begin{equation}
\label{eq:e_min}
e_{\rm min}\approx \frac{x_h}{Q} \sqrt{1-1/(4\beta^2)}  \ .
\end{equation}
The scalar field profiles of scalar clouds for some values of $e$ between $e_{\rm max}$ and $e_{\rm min}$ are shown in Fig. \ref{data_rn} (left) for $x_h = 0.15$ and $Q = 0.09$. For this RN black holes, which has $60\%$ of its maximal possible charge, we get from the formulae above that $e_{\rm max}=5/3\approx 1.67$ and $e_{\rm min}\approx 5/9 \approx 0.56$. Note that our numerical results agree very well with these values, we 
find $e_{\rm min}\approx 0.6$ numerically, see Fig. \ref{data_rn} (right), where we give $\psi(x_h)=\psi_h$, 
$\psi'\vert_{x=x_h}$ as well as $\mu_{{\rm eff},\infty}$ in function of $e$. Obviously, both the scalar field and its derivative on the horizon tend to  zero for $e\rightarrow e_{\rm max}$. That one follows from the other can, in fact, be understood when remembering the regularity condition
on the horizon, see (\ref{eq:horizon_condition}). Approaching $e_{\rm min}$, $\psi_h$ tends to the value of 
the local minimum of the ``bare'' potential $U(\psi)$ (remember that $v(x_h)=0$), which for our choice of parameters is $\psi\approx 1.286$.
This implies -- again using  (\ref{eq:horizon_condition}) -- that $\psi'\vert_{x=x_h}$ should approach zero. 
We find that  $\mu_{{\rm eff},\infty}$ tends to zero at $e\rightarrow e_{\rm max}$ and
to $\approx 1/(2\beta)\approx 0.94$ at $e\rightarrow e_{\rm min}$. 
As our reasoning above suggests, both $e_{\rm min}$ and $e_{\rm max}$ decrease for increasing charge $Q$ at fixed horizon radius. Close to the extremely
charged RN solution with $Q\rightarrow x_h$ we find that $e_{\rm max}\rightarrow 1$ and $e_{\rm min}\rightarrow 1/3$, which again agrees
very well with our results. 
In other words, two black holes with equal event horizon radius, but different charge differ in the required coupling strength
$e$ to allow for the existence of $Q$-clouds. The larger the charge of the black hole, the smaller
the gauge coupling can be chosen for $Q$-clouds to exist. This is related to the electromagnetic
repulsion of the cloud which consists of scalar boson that are coupled to the electromagnetic
field of the RN black hole. However, note that extremal RN black holes with $x_h=Q$ cannot have regular scalar clouds, as (\ref{eq:horizon_condition}) clearly demonstrates. 
The mass $M_Q$ (see (\ref{eq:mq})) of the cloud becomes
\begin{equation} 
M_Q=\frac{Q^2}{x_h}  - 2 \int\limits_{x_h}^{\infty} {\rm d}x \ x^2 \ U(\psi)     \ ,
\end{equation}
i.e. is determined by the properties of the black hole and the integral of the scalar potential energy, respectively. 
Also note that a constant, non-vanishing value of the scalar field leads to a decrease of $M_Q$ when the interval
in $x$ on which $\psi$ is constant (and non-zero) increases. 
\subsection{Schwarzschild black holes}
\begin{figure}[ht!]
\hspace{-1cm}
\includegraphics[width=8.5cm]{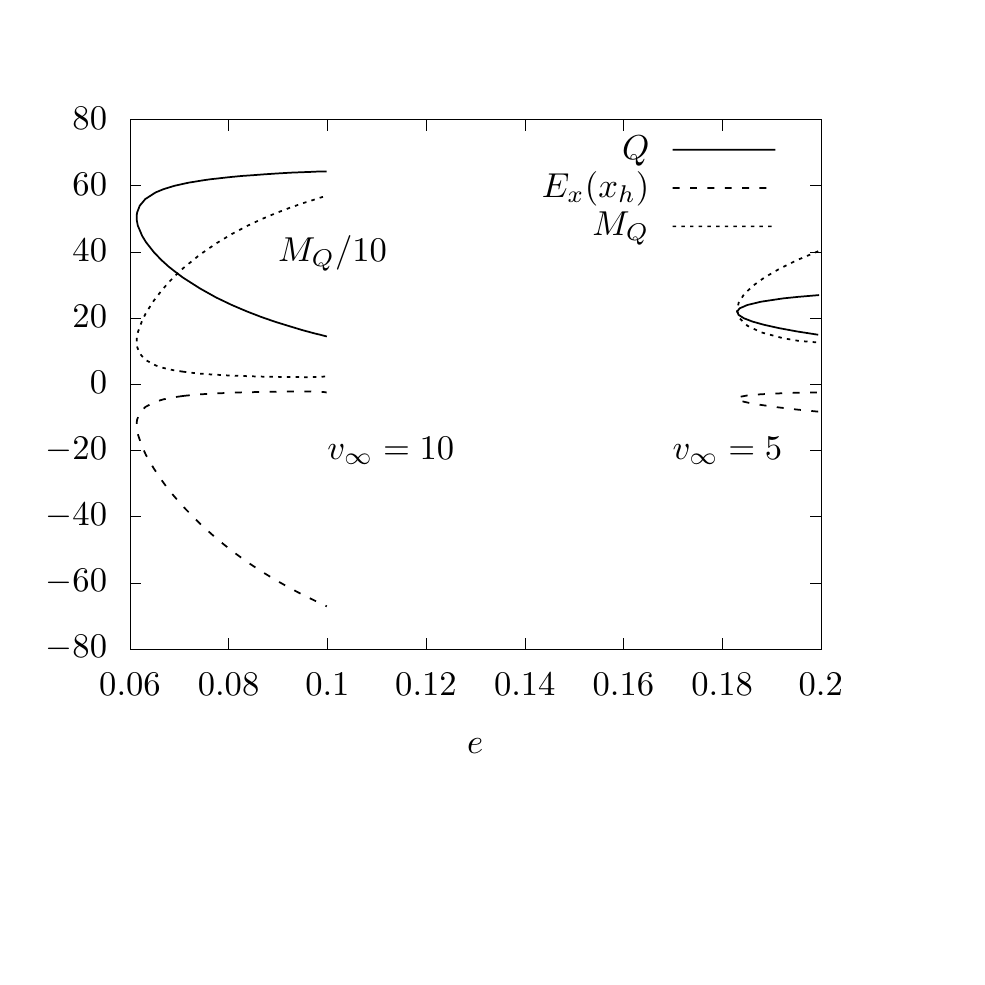}
\includegraphics[width=8.5cm]{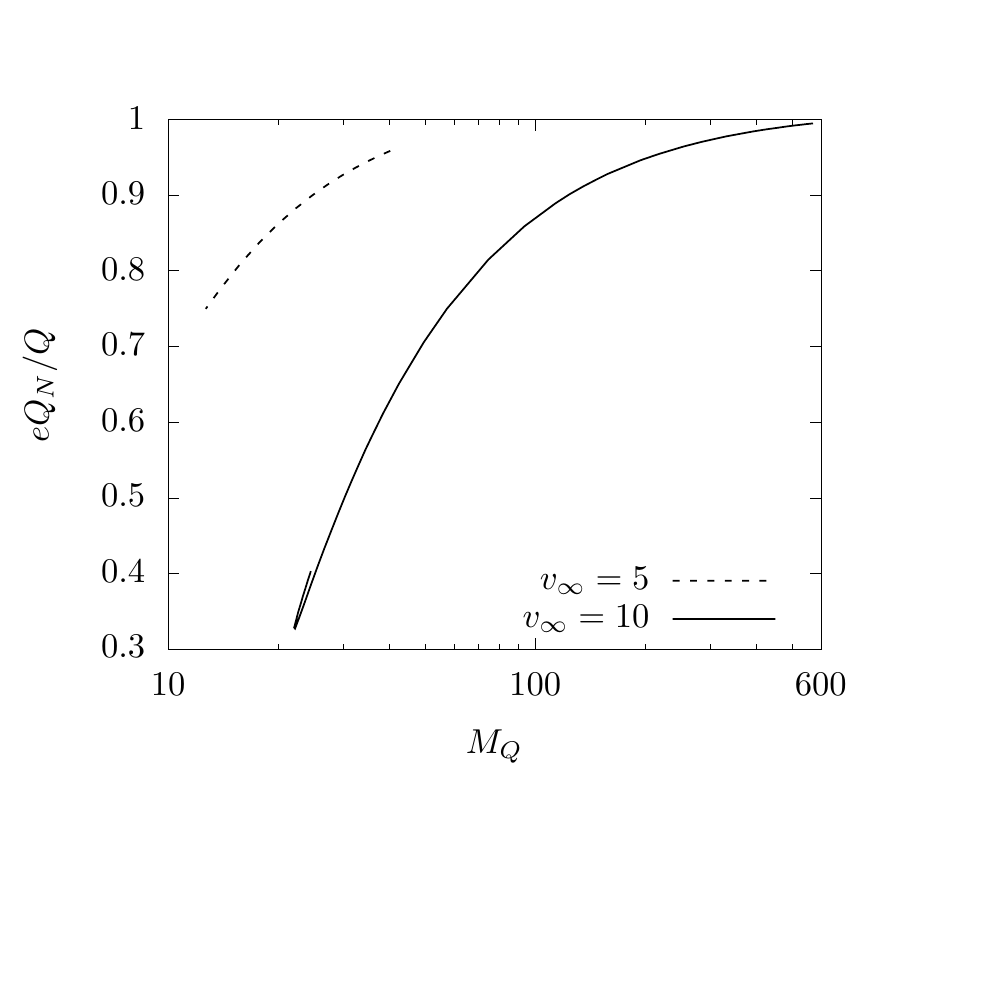}
\vspace{-2.5cm}
\caption{{\it Left}~:We show the dependence of the electric charge $Q$, the radial electric field at the horizon $E_x(x_h)=-v'(x_h)$ as well as the mass $M_Q$ of charged $Q$-clouds around Schwarzschild black holes in dependence of the gauge coupling $e$ for two values of $v_{\infty}$.
{\it Right}~: We show the ratio of $eQ_N$, i.e. the total charge of the $Q_N$ individual bosonic
particles making up the cloud, and the total electric charge $Q$ as function of the mass $M_Q$ for
two values of $v_{\infty}$.}
\label{fig:data_schw}
\end{figure}

In this section, we consider the equations for the gauge and scalar field, respectively, in the background of a Schwarzschild black hole, i.e. we set $\alpha^2=0$ in (\ref{eq:m_sigma}) such that the gravitational field is not
sourced by any matter field. The result is the Schwarzschild solution with
$N(x)=1-2M/x$, $\sigma\equiv 1$ with an event horizon at radius $x=x_h=2M$. 
In contrast to the RN case, the gauge field and scalar field equations are now coupled. The appropriate boundary conditions for the equations (\ref{eq:V}) and (\ref{eq:psi}) are~: 
\begin{equation}
   \psi'\vert_{x=x_h}  = \frac{x_h}{2} \frac{dU}{d\psi}\biggr\vert_{\psi=\psi_h}  \ \ , \ \ v(x_h) = 0 \ \ , \ \ \psi(x \to \infty) = 0 \ \ , \ \ v(x \to \infty) = v_{\infty} \ .
\end{equation}
We have then fixed the strength of the attractive self-interaction, $\beta^2$, the value of the
gauge field at infinity, $v_{\infty}$, and the radius of the black hole horizon, $x_h$, and studied the
dependence of the solutions on the strength of the gauge coupling $e$, i.e. the coupling between the
gauge field and the scalar field. Charged $Q$-clouds around Schwarzschild black holes have been
studied before in \cite{Herdeiro:2020xmb}, however, we find that not one, but {\bf two} solutions
exist for one choice of parameters $\beta^2$, $v_{\infty}$, $x_h$ and $e$. These two solutions
differ in the size of the scalar cloud and consequently also in their energy $M_Q$ , 
the electric charge $Q$ and the value of the radial electric field $E_x=-v'$ on the horizon. 
This is shown in Fig. \ref{fig:data_schw} for $x_h = 0.15$, $\beta^2 = 9/32$ and two different
values of $v_{\infty}$. Both branches exist on a finite interval of the gauge coupling $e$
with maximal coupling strength given by $1 - e_{\rm max}^2 v_{\infty}^2=0$ (see (\ref{eq:fall_off_scalar})). In agreement
with this reasoning, we find $e_{\rm max}=0.1$ (resp. $e_{\rm max}=0.2$) for $v_{\infty}=10$ ($v_{\infty}=5$). At the minimal value of $e$, $e_{\rm min}$, the two branches of solutions join. 
The value of $e_{\rm min}$ can only be determined numerically and we find that
$e_{\rm min}\approx 0.061$ (resp. $e_{\rm min}\approx 0.183$) for $v_{\infty}=10$ ($v_{\infty}=5$).
To state it differently~: the larger $v_{\infty}$, the smaller we have to choose the value of the gauge coupling in order to allow for charged $Q$-clouds to exist. 
On the other hand, the larger $v_{\infty}$, the larger the interval $e_{\rm max}-e_{\rm min}$ on which charged $Q$-clouds exist. For $v_{\infty}\approx 4$, this interval shrinks to zero, i.e.
scalar clouds on Schwarzschild black holes do no longer exist if the potential
difference between the horizon and infinity is too small.
At fixed gauge coupling $e$, the clouds on the second branch have higher mass $M_Q$ and higher
electric charge $Q$ than those of the first, already known branch. Moreover, the radial electric field on the horizon, $E_x(x_h)$, is smaller in absolute value.  It is also interesting to consider the
ratio between the total electric charge of the $Q_N$ individual bosons making up the cloud,
$eQ_N$, and the total electric charge $Q$ of the solution. This ratio is shown in dependence
of the mass $M_Q$ in Fig. \ref{fig:data_schw} (right). We find for both values of $v_{\infty}$ studied
that the increase of the ratio leads to an increase of the mass $M_Q$. Remembering that
the total electric charge $Q$ is a sum of the horizon electric charge and the charge in the $Q_N$
bosons in the cloud, we conclude that transfer of electric charge from the horizon to the scalar cloud increases the mass of the solution. Moreover, there are limits (depending on $v_{\infty}$) how much
transfer can be done. For $v_{\infty}=5$, we need at least $\approx 75 \%$ of the electric charge $Q$ to be in the cloud and are allowed maximally $\approx 96 \%$. For $v_{\infty}=10$, the lower
bound is $\approx 32\%$, while the upper bound is $\approx 99 \%$, i.e. very close to the 
limit, where all electric charge is in the cloud. Remember that only globally regular solutions, so-called boson stars, have $eQ_N=Q$. 
\\

\begin{figure}[ht!]
\vspace{-1cm}
\begin{center}
\includegraphics[width=10cm]{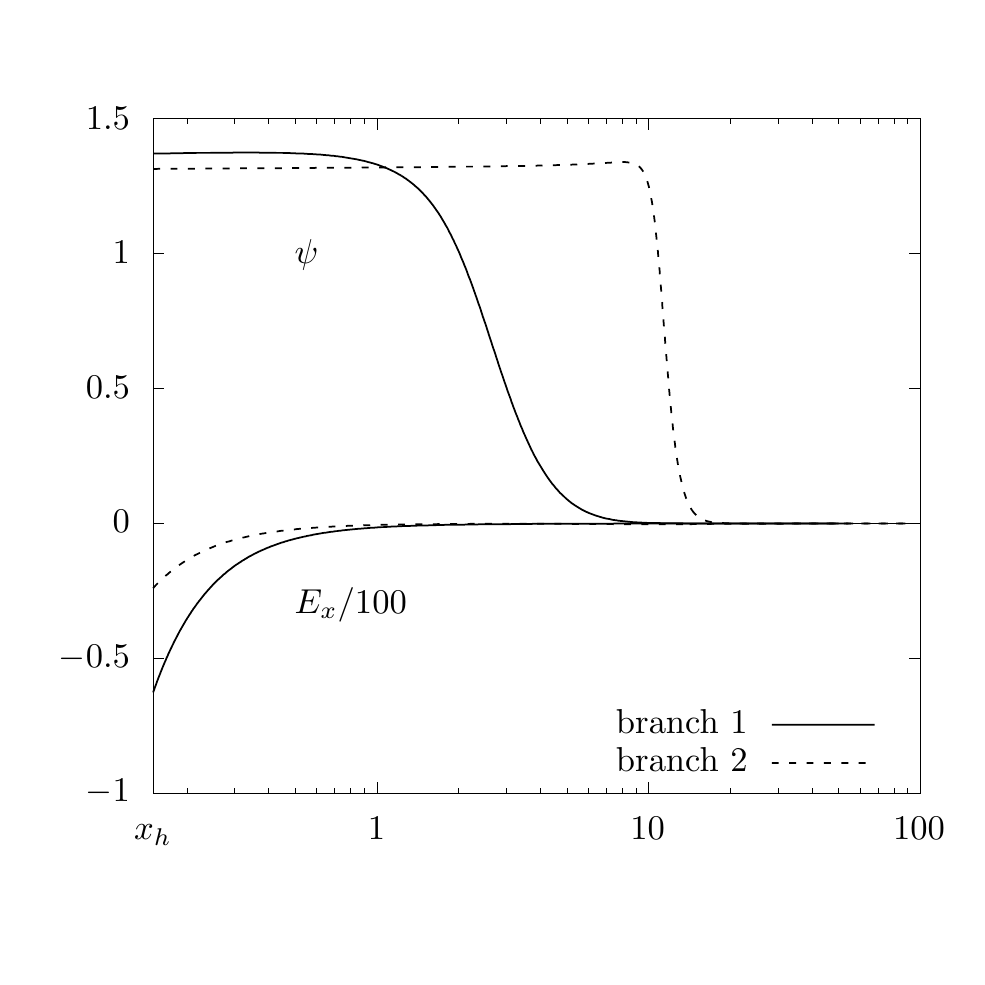}
\end{center}
\vspace{-1,5cm}
\caption{We compare the scalar ($\psi$) and radial electric field ($E_x$) profiles of the two $Q$-cloud solutions in a Schwarzschild space-time with $x_h=0.15$ for $v_{\infty} =10$ and $e=0.08$.}
\label{fig:branches_schw}
\end{figure}

How the transfer of charge to the cloud is possible becomes  clear when considering the
two solutions available at the same
values of the couplings. We have plotted the profiles of the radial electric field $E_x$ as well as the scalar field $\psi$ of the two solutions available at $v_{\infty}=10$ and $e=0.08$
in Fig.\ref{fig:branches_schw}. The solution on the second branch
has a practically constant, non-vanishing scalar field outside the horizon, i.e. is non-vanishing
for a much larger interval of the radial coordinate than the corresponding solution on the first branch. Hence, the extend of the scalar cloud is larger for solutions on the second branch. Note also that -- very similar to what we have observed for clouds on RN black holes --
the value of the scalar field on the horizon approaches the local minimum of the ``bare'' potential $\psi_{min,2}\approx 1.286$. 
This value extends to increasingly larger values of $x$ along the branches and shows a sudden drop to the global minimum $\psi_{min,1}=0$.
In order words~: the scalar clouds possess an interior (containing the black hole), in which the scalar field
is trapped in the ``false vacuum'' of the potential  $\psi_{min,2}$. 
The explanation of the existence of the second branch and the behaviour of the scalar field becomes
clear when remembering that $e=g/\sqrt{\lambda}$. On the first branch of solutions, varying $e$ can be interpreted as
varying the gauge coupling $g$ and fixing $\lambda$. This can be done down to a minimal value of $e$, as discussed above.
Now, increasing $e$ from this minimal value on the second branch can be interpreted as keeping $g$ fixed
and decreasing $\lambda$. 

\begin{figure}[ht!]
\begin{center}
\includegraphics[width=10cm]{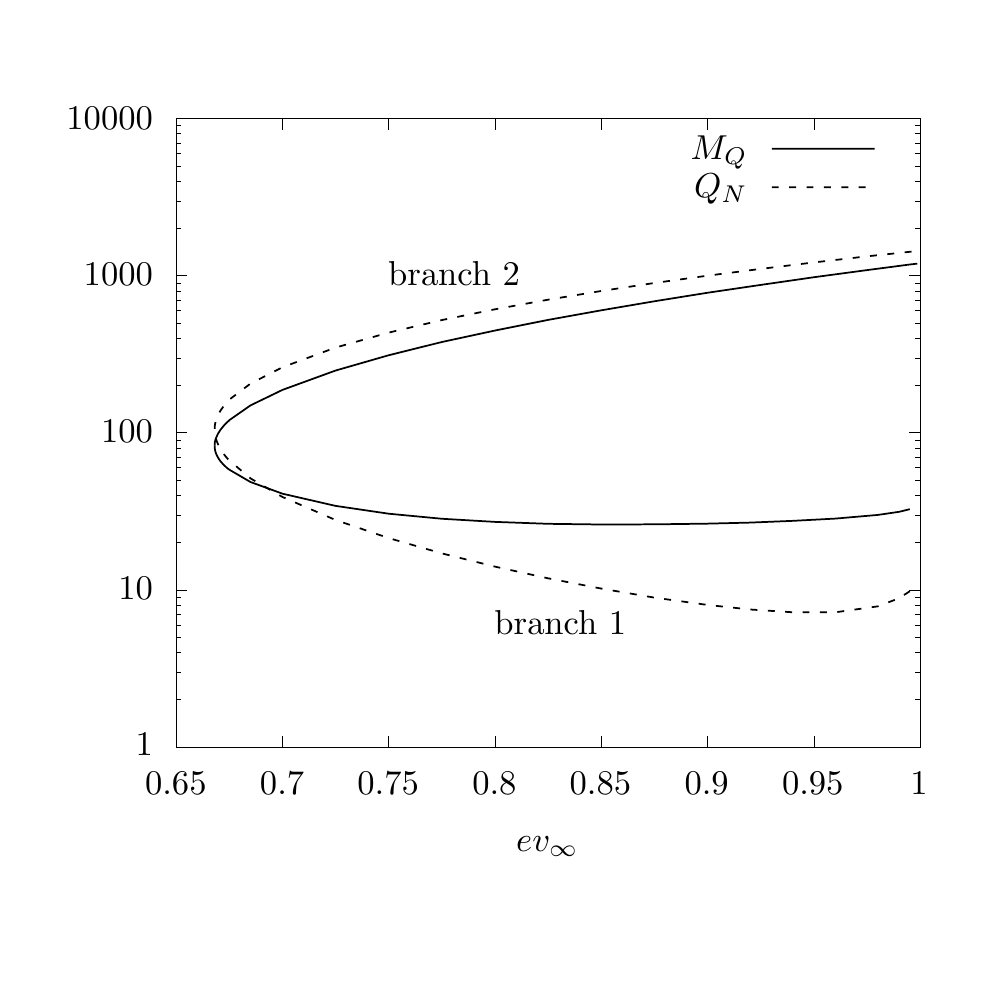}
\end{center}
\vspace{-1.5cm}
\caption{We show the mass $M_Q$ and the Noether charge $Q_N$ of the two branches
existing for scalar clouds on a Schwarzschild black hole with $x_h=0.15$ for $e=0.08$ and in dependence on $ev_{\infty}$. }
\label{fig:branches_schw2}
\end{figure}

The fact that two different solutions exist for the same choice of coupling constants
can also be observed when fixing $e$ and varying $v_{\infty}$. 
This is shown in Fig. \ref{fig:branches_schw2} for $e=0.08$ and $x_h=0.15$, where
we give the mass $M_Q$ and the Noether charge $Q_N$ in dependence of $ev_{\infty}$. 
Again, both branches extend all the way back to $ev_{\infty}=1$, where the effective
mass of the scalar field $1-e^2 v_{\infty}^2$ becomes zero. 
Interestingly, we find that the solutions on the first branch have $M_Q > Q_N$, while
the ones on the second branch have $M_Q < Q_N$. 
In fact, we can approximately calculate the value of $M_Q$ for the solutions on the second branch with pronounced
``hard wall''. First, assume that $\psi\equiv \psi_0$ for $x\in[x_h:\tilde{x}]$ and $\psi\equiv 0$ for $x\in [\tilde{x}:\infty [$.
Inserting this into the gauge field equation (\ref{eq:V}), this gives
\begin{equation}
\label{eq:Vscreened}
v(x)\sim \begin{cases} v_{\infty} - \frac{Q}{x}\exp(-\sqrt{2} e\psi_0 x) & \ \ {\rm for } \ \  x \in [x_h:\tilde{x}] \\
\\
v_{\infty} - \frac{Q}{x} & \ \ {\rm for } \ \  x \in [\tilde{x}:\infty [ \end{cases}  \ .
\end{equation}
For $x\in [\tilde{x}:\infty [$, this is the standard electric potential of a point charge, for $x \in [x_h:\tilde{x}]$, this is
the electric potential of a {\it screened point charge}.  Now using the form of $\psi$ and $v$ and inserting this into (\ref{eq:mq}), we find that
\begin{multline}
M_Q  =  2 Q v_{\infty} - \frac{Q}{\tilde{x}} - \frac{Q^2 e \psi_0}{\sqrt{2}} \left[\exp(-2\sqrt{2} e\psi_0 x_h) - 
\exp(-2\sqrt{2} e\psi_0 \tilde{x})\right] \nonumber \\
 -   Q^2 \left[\frac{\exp(-2\sqrt{2} e\psi_0 x_h)}{x_h} - 
\frac{\exp(-2\sqrt{2} e\psi_0 \tilde{x})}{\tilde{x}}\right] - \frac{2}{3} U(\psi_0) \left(\tilde{x}^3 - x_h^3\right) \ .
\end{multline}
As a final remark, let us state that all charged $Q$-clouds on Schwarzschild black holes possess charge  larger than their charged black hole counterparts with the same horizon radius. We find for all solutions that $Q > x_h$ or equivalently $v_{\infty}  > v_{\infty, {\rm RN}} \equiv Q/x_h$. 
\section{Back-reaction of $Q$-clouds on space-time}
\begin{figure}[ht!]
\hspace{-1cm}
\includegraphics[width=8.5cm]{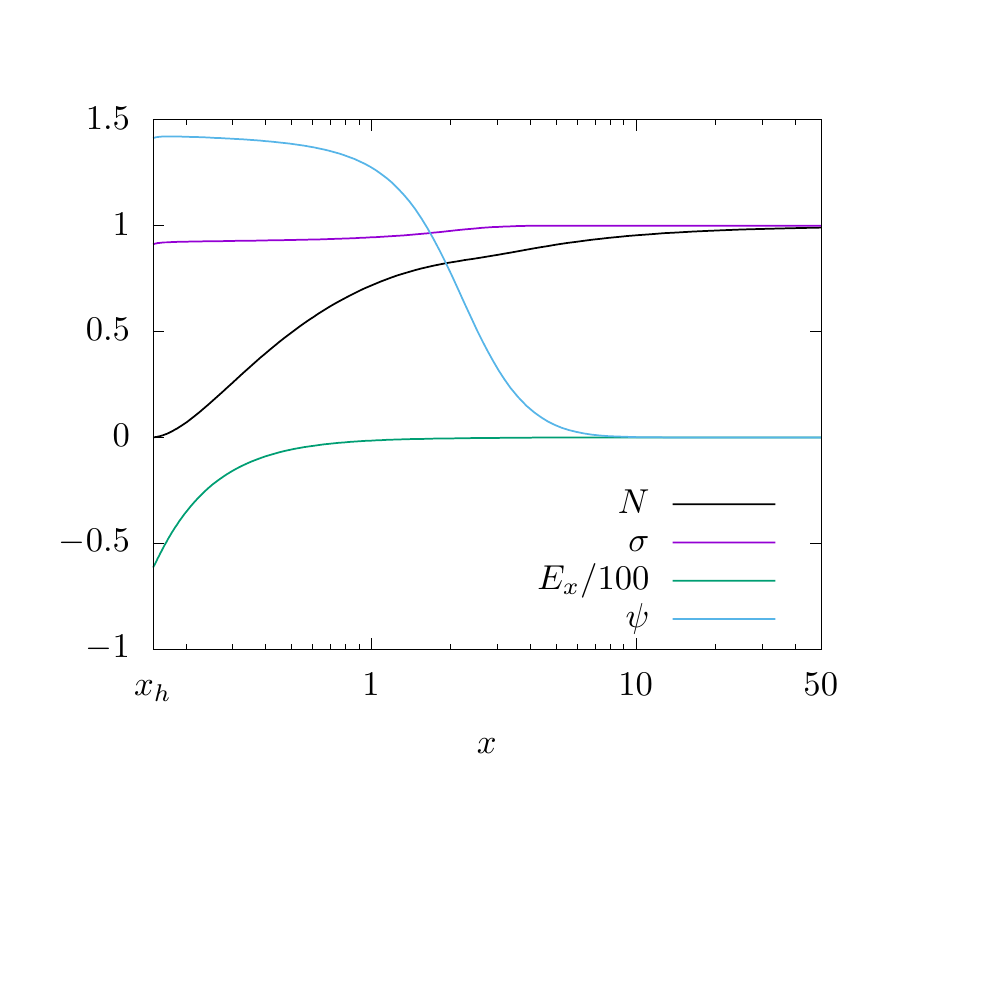}
\includegraphics[width=8.5cm]{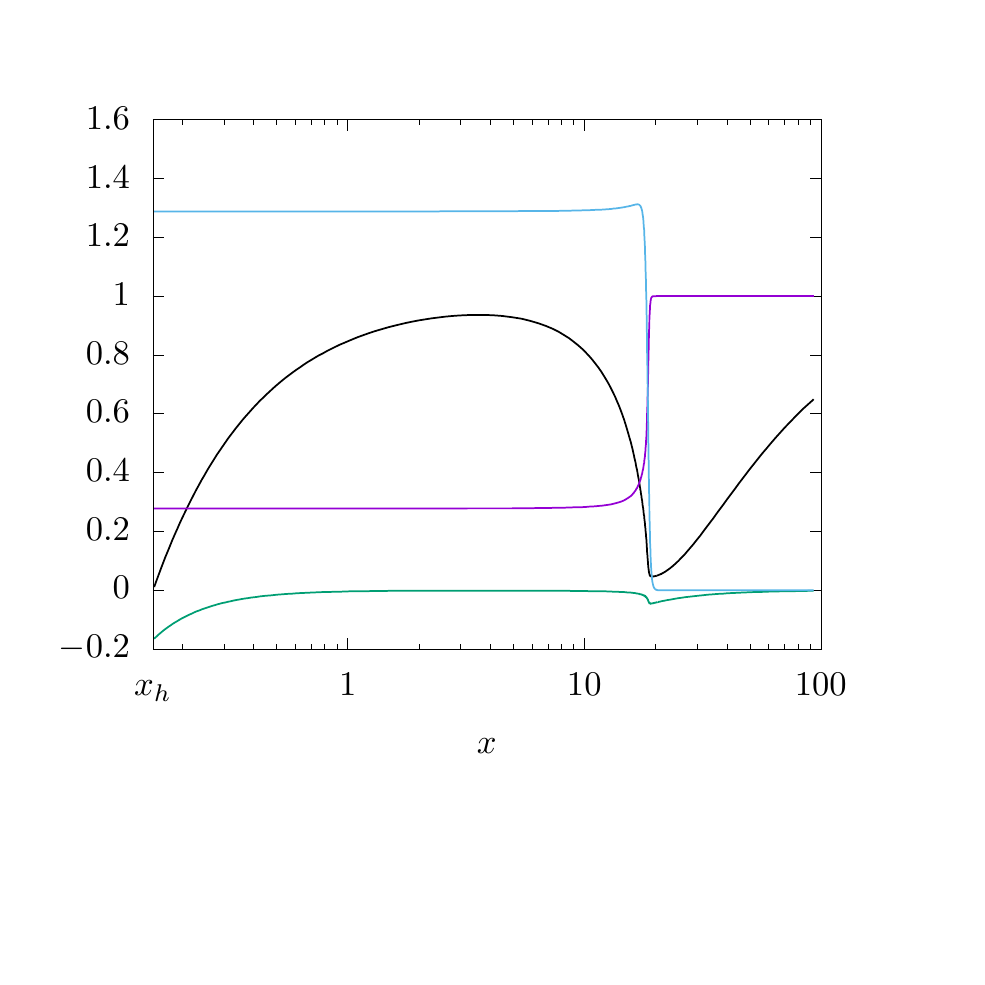}
\vspace{-2cm}
\caption{{\it Left}~: We shown the profiles of the metric functions $N$, $\sigma$, the radial electric field $E_x$ and the scalar field $\psi$ for a scalar cloud on the first branch of solutions
for $x_h=0.15$, $v_{\infty} =10$, $e = 0.08$ and at maximal possible back-reaction $\alpha_{cr,1}^2\approx 0.0095$. {\it Right}~: Same as left, but for a scalar cloud on the second
branch of solutions close to the maximal possible back-reaction $\alpha_{cr,2}^2\approx 0.0124$. }
\label{fig:gravity}
\end{figure}

The question is then how these scalar clouds, once formed, back-react on the space-time.
This is what we will discuss here. We have solved the system of coupled
differential equations (\ref{eq:m_sigma}), (\ref{eq:V}) and (\ref{eq:psi}) subject to the appropriate
boundary conditions.
Considering the scalar cloud in the background of a Schwarzschild black hole, we have then
increased the back-reaction of this cloud onto the space-time by increasing the parameter
$\alpha^2$ from zero. As mentioned above, two solutions exist for the same values of the parameters and our results indicate that, when increasing the back-reaction, these solutions behave very differently in the strong gravity regime. 
Increasing the gravitational back-reaction for the scalar cloud with lower mass at fixed gauge coupling, we find that the temperature $T_H$
of the black hole decreases and approaches zero at a critical value of $\alpha$, $\alpha_{1,cr}$. This suggests that
the limiting solution is an extremal black hole. Indeed, we observe that the metric function $N(x)$ develops a double zero at $x=x_h$, while
$\sigma(x_h)$ stays perfectly finite. This can be seen in Fig. \ref{fig:gravity} (left), where we plot the metric
functions $N(x)$, $\sigma(x)$, the radial electric field $E_x(x)$ as well as the scalar field $\psi(x)$ for $x_h=0.15$, $e=0.08$, $v_{\infty}=10$ and
the value of $\alpha$ close to the maximal possible value, $\alpha_{1,cr}^2\approx 0.0095$. Note that this extremal black hole, however,
does not possess regular scalar hair as the derivative of the scalar field, $\psi'$ diverges on the extremal horizon. That this should
be so (and is confirmed by our numerics) is obvious when remembering the boundary conditions (\ref{eq:bc_horizon}). 
On the other hand, when increasing $\alpha$ for a solution on the second branch,
the metric function $N(x)$ develops a local
minimum at $x\approx \tilde{x} > x_h$ which corresponds to the value of $x$ where $\psi$ drops from its roughly constant and non-vanishing value to zero. At the maximal possible back-reaction $\alpha=\alpha_{cr,2}$ the value of this minimum drops to zero and
the solution forms a second horizon
at $x_{h,2}\approx \tilde{x}$. For 
$x \in [x_{h,2}: \infty]$ the solution possesses no scalar field, $\psi\equiv 0$, and the space-time corresponds
to an extremal Reissner-Nordstr\"om solution. 
This is shown in Fig. \ref{fig:gravity} (right). In the region $x\in [x_h, x_{h,2}]$, the 
solution possesses a constant, but non-vanishing scalar field $\psi\equiv \psi_0$.  Hence,
the scalar potential sources the space-time in a non-trivial way. 
Using the fact that our numerics indicates $\sigma\equiv \sigma_0\neq 1$ for $x\in [x_h:\tilde{x}]$
we can combine the gravity equations (\ref{eq:m_sigma})  to find
\begin{equation}
N(x)= 1 - \frac{2M}{x} - \frac{\alpha^2}{x\sigma_0^2}\int {\rm d} x \left(x v'\right)^2 - \frac{2\alpha^2}{3} U(\psi_0) x^2  \  \ \ {\rm for} \ \ 
x\in [x_h:\tilde{x}] \ .
\end{equation}
Using the gauge field $v$ for a screened charge $Q$ (see (\ref{eq:Vscreened})) the integration leads to~:
\begin{equation}
N(x)=1-\frac{2M}{x} + \left[\frac{\alpha^2 Q^2}{\sigma_0^2 x^2} + \frac{\sqrt{2} \alpha^2 Q^2 e \psi_0}{2x}\right] \exp\left(-2\sqrt{2} e \psi_0 x\right) - \frac{2\alpha^2}{3}  U(\psi_0) x^2  \  \ \ {\rm for} \ \ 
x\in [x_h:\tilde{x}] \ .
\end{equation}
which becomes the RN solution for $\psi_0=0$, $\sigma_0=1$. The metric function describes
a space-time with mass $M$ and a screened electric charge $Q$ as well as a cosmological constant term with $\Lambda=2\alpha^2 U(\psi_0)$. The space-time is hence a RN-de Sitter (RNdS)
solution with screened electric charge. For large $x$ this space-time has a cosmological horizon
at $x_{\Lambda}=1/(\alpha\sqrt{2 U(\psi_0)})$. 
For the solution presented in Fig. \ref{fig:gravity} (right), we find that $\alpha_{cr,2} \approx 0.0124$,
and $\psi_0\approx 1.28$ such that $x_{\Lambda}\approx 14.5$. Of course, this does not
agree with the result we find exactly, since also the mass term 
and the screened charge have to be taken into account when computing the horizon,  but it gives a good approximation to our result. So, the extremal horizon for the exterior solution
corresponds to a cosmological horizon for the interior solution. In other words, the limiting
solution is an inflating black hole with carries a screened charge and looks like an extremal Reissner-Nordstr\"om 
black hole from the outside. 
\vspace{1cm}
\section{Conclusions}
In this paper, we have discussed scalar clouds on static, spherically symmetric black holes as well as the back-reaction
of these clouds onto the space-time. 
Studies of these solutions have been done before in \cite{Hong:2019mcj, Herdeiro:2020xmb, Hong:2020miv}, however, the strong gravity
limit had not been discussed previously. Investigating the charged clouds on a fixed Schwarzschild background, we
find a new branch of solutions that had not been noticed previously. In contrast to scalar clouds on RN black holes, where the cloud 
disperses to spatial infinity at minimal coupling to the electric field of the black hole, the extend of charged scalar clouds on Schwarzschild black holes is limited and forms a ``hard wall'' on which the energy density associated to the scalar field abruptly drops to zero. 
We also find that for these type of solutions we can transfer practically all electric charge to the scalar cloud when choosing the potential
difference between the horizon and infinity large enough. 
When investigating the back-reaction of these clouds onto the space-time, we observe that at the ``hard wall'' a horizon forms.
The scalar field has a value close to the second minimum of the scalar potential, i.e. is trapped inside the ``false vacuum'' of the
model. This generates a huge amount of scalar potential energy that back-reacts onto the space-time in the form of
a positive cosmological constant. For strong enough back-reaction, this leads to the formation of an extremal black hole.
This phenomenon is reminiscent to that of an inflating monopole in the context of topological inflation \cite{Linde:1994hy,Vilenkin:1994pv}, 
where the inside of the monopole core is trapped inside the false vacuum of the potential. The energy associated to this false vacuum leads to the collapse of the monopole
to a magnetically charged RN solutions at sufficiently strong back-reaction  Although we have discussed a
scalar field theory that possesses an unbroken U(1) symmetry (in contrast to the monopole case), the scalar field gets trapped
inside the ``false vacuum'' of the potential and leads to an inflating black hole solution. 
It would certainly be interesting to see whether such a phenomenon could also be observed in the case of rotating black holes and
whether our results have any relevance to the theory of inflation. 

\vspace{1cm}
{\bf Acknowledgements} Y. B. would like to thank Eugen Radu for discussions in the initial stages of this paper.
B. H. would like to thank FAPESP for financial support under grant {\it 2019/01511-5}
as well as the DFG Research Training Group 1620 {\it Models of Gravity} for financial support. 
B.H. would also like to thank Jose J. Blanco-Pillado for bringing the papers  \cite{Linde:1994hy,Vilenkin:1994pv}
to her attention. 
\clearpage

\end{document}